\documentclass[prx,twocolumn,superscriptaddress,longbibliography]{revtex4-2}
\usepackage{makeidx}
\usepackage[utf8]{inputenc}
\usepackage{graphicx}
\usepackage{amsmath}
\usepackage{mathrsfs}
\usepackage{graphicx}
\usepackage{braket}
\usepackage[subrefformat=parens,labelformat=parens,caption=false]{subfig}
\usepackage{amsfonts}
\usepackage{dcolumn}
\usepackage{bm}
\usepackage{bbm}
\usepackage{color}
\usepackage[dvipsnames]{xcolor}
\usepackage{esint}
\usepackage{lineno}
\usepackage{verbatim}
\usepackage{cancel}
\usepackage{times} 
\usepackage[breaklinks,
            colorlinks,
            urlcolor=blue,
            linkcolor=blue,
            anchorcolor=blue,
            citecolor=blue]{hyperref}
           
\usepackage{soul}
\usepackage{amssymb}

\definecolor{THc}{rgb}{0.9,0.3,0.2}

\newcommand{\EE}[1]{\underset{#1}{\mathbb{E}}}

\renewcommand{\eqref}[1]{Eq.~(\ref{#1})}

\newcommand{\sectionMain}[1]{
\let\oldaddcontentsline\addcontentsline
\renewcommand{\addcontentsline}[3]{}
\section{#1}
\let\addcontentsline\oldaddcontentsline
}

\allowdisplaybreaks

\begin{document}

\title{Rigorous noise reduction with quantum autoencoders}

\author{Wai-Keong Mok*}
\affiliation{California Institute of Technology, Pasadena, CA 91125, USA}
\affiliation{Centre for Quantum Technologies, National University of Singapore, 3 Science Drive 2, Singapore 117543}
\thanks{These authors contributed equally to this work}

\author{Hui Zhang*}
\affiliation{Institute of Quantum Technologies (IQT), The Hong Kong Polytechnic University, Hong Kong}
\affiliation{Quantum Science and Engineering Centre (QSec), Nanyang Technological University, Singapore}
\thanks{These authors contributed equally to this work}
\author{Tobias Haug*}
\affiliation{QOLS, Blackett Laboratory, Imperial College London SW7 2AZ, UK}
\thanks{These authors contributed equally to this work}
\author{Xianshu Luo}
\affiliation{Advanced Micro Foundry, 11 Science Park Rd, Singapore}

\author{Guo-Qiang Lo}
\affiliation{Advanced Micro Foundry, 11 Science Park Rd, Singapore}

\author{Hong Cai}
\affiliation{Institute of Microelectronics, A*STAR (Agency for Science, Technology and Research), Singapore}

\author{M. S. Kim}
\affiliation{QOLS, Blackett Laboratory, Imperial College London SW7 2AZ, UK}

\author{Ai Qun Liu} \affiliation{Institute of Quantum Technologies (IQT), The Hong Kong Polytechnic University, Hong Kong}
\affiliation{Quantum Science and Engineering Centre (QSec), Nanyang Technological University, Singapore}

\author{Leong-Chuan Kwek}
\affiliation{Centre for Quantum Technologies, National University of Singapore, 3 Science Drive 2, Singapore 117543}
\affiliation{MajuLab, CNRS-UNS-NUS-NTU International Joint Research Unit, Singapore UMI 3654, Singapore}
\affiliation{National Institute of Education, Nanyang Technological University, Singapore 637616, Singapore}
\affiliation{Quantum Science and Engineering Centre (QSec), Nanyang Technological University, Singapore}

\begin{abstract}
Reducing noise in quantum systems is a major challenge towards the application of quantum technologies. 
Here, we propose and demonstrate a scheme to reduce noise using a quantum autoencoder with rigorous performance guarantees. The quantum autoencoder learns to compresses noisy quantum states into a latent subspace and removes noise via projective measurements.
We find various noise models where we can perfectly reconstruct the original state even for high noise levels. 
We apply the autoencoder to cool thermal states to the ground state and reduce the cost of magic state distillation by several orders of magnitude.
Our autoencoder can be implemented using only unitary transformations without ancillas, making it immediately compatible with the state of the art.
We experimentally demonstrate our methods to reduce noise in a photonic integrated circuit. 
Our results can be directly applied to make quantum technologies more robust to noise. 
\end{abstract}

\maketitle

\section*{Introduction}
Quantum technologies offer potential advantages in quantum computing~\cite{montanaro2016quantum}, quantum communication~\cite{pirandola2020advances} and metrology~\cite{giovannetti2011advances}. However, quantum systems are brittle by nature, and noise due to the environment and imperfect control over the quantum system negatively impacts the capabilities of quantum devices. Thus, techniques to remove or reduce noise is the key challenge that needs to be addressed for quantum technologies to be successful~\cite{suter2016colloquium,bharti2022noisy}.
To this end, a wide range of noise reduction techniques have been developed.

In the context of fault-tolerant quantum computers~\cite{gottesman2010introduction}, magic state distillation (MSD) requires multiple copies of a noisy state to create one state with reduced noise. MSD is the most expensive process required to run fault-tolerant quantum computers~\cite{bravyi2012magic} and it is imperative to substantially reduce the cost of MSD to make early fault-tolerant quantum computers practically viable~\cite{campbell2017roads,suzuki2022quantum,krishna2019towards,hastings2018distillation,jones2013multilevel,campbell2017unifying,haah2017magic}. 

An alternative path to reduce noise are quantum autoencoders.
Quantum autoencoders transform quantum states into a smaller subspace that contains the essential features of the state, while discarding redundant features~\cite{romero2017quantum,wan2017quantum,lamata2018quantum,du2021exploring,bravo2021quantum,anand2022quantum,zhang2022resource,liu2023information}. Quantum autoencoders are amenable to noisy quantum devices~\cite{bharti2022noisy} which makes quantum autoencoders particularly useful for enhancing quantum technologies in experiments~\cite{pepper2019experimental,zhang2022resource,huang2020realization,zhou2022preserving,ding2019experimental}. 
Experiments have demonstrated loss compression of quantum data in bulky optics systems~\cite{pepper2019experimental,huang2020realization} and the use of autoencoders for compression-assisted teleportation of high-dimensional quantum states in integrated photonic chips~\cite{zhang2022resource}.

Recently, quantum autoencoders have been  proposed to reduce noise~\cite{bondarenko2020quantum,zhang2021generic,achache2020denoising,cao2021noise,locher2023quantum,pazem2023error,tran2023variational}. 
One variant reduces noise by transferring the noisy part of the state into ancillas and tracing out the ancilla. This approach allows for deterministic noise removal, however it commonly requires deep circuits and many ancilla qubits~\cite{bondarenko2020quantum}. 
Alternatively, projective measurements and post-selection can reduce noise with low resource requirements and without ancillas~\cite{zhang2021generic}. However, existing proposals provide mainly numerical evidence for the performance in practical applications. A quantum autoencoder to denoise quantum states is yet to be experimentally demonstrated.

Here, we experimentally implement a photonic chip integrated autoencoder to reduce the noise of quantum states with rigorous performance guarantees.
Our scheme compresses quantum states into a latent subspace and removes noise by projective measurements and post-selection on successful outcomes. We train the autoencoder either in an unsupervised manner by minimizing measurement probabilities of noisy input states (\emph{population training}), or maximizing the fidelity in respect to a reference state (\emph{fidelity training}).
We analytically study the performance of the protocol and provide rigorous bounds on the denoising fidelity. For various noise models such as perturbation by a fixed state, depolarizing noise or thermal states, we find that our protocol can perfectly recover the noise-free state. 
We also apply it to cool a thermal state to the ground state. Further, we show that our protocol can decrease the cost of magic state distillation by several orders of magnitudes. Remarkably, this allows for successful distillation at high levels of noise where the conventional protocol fails.
The protocol is experimentally realized on an integrated photonic chip, which is scalable and energy-efficient. 
Our work demonstrates a practical method to reduce noise for immediate applications in quantum technologies.

\begin{figure*}
\centering
\subfloat{%
  \includegraphics[width=0.97\linewidth]{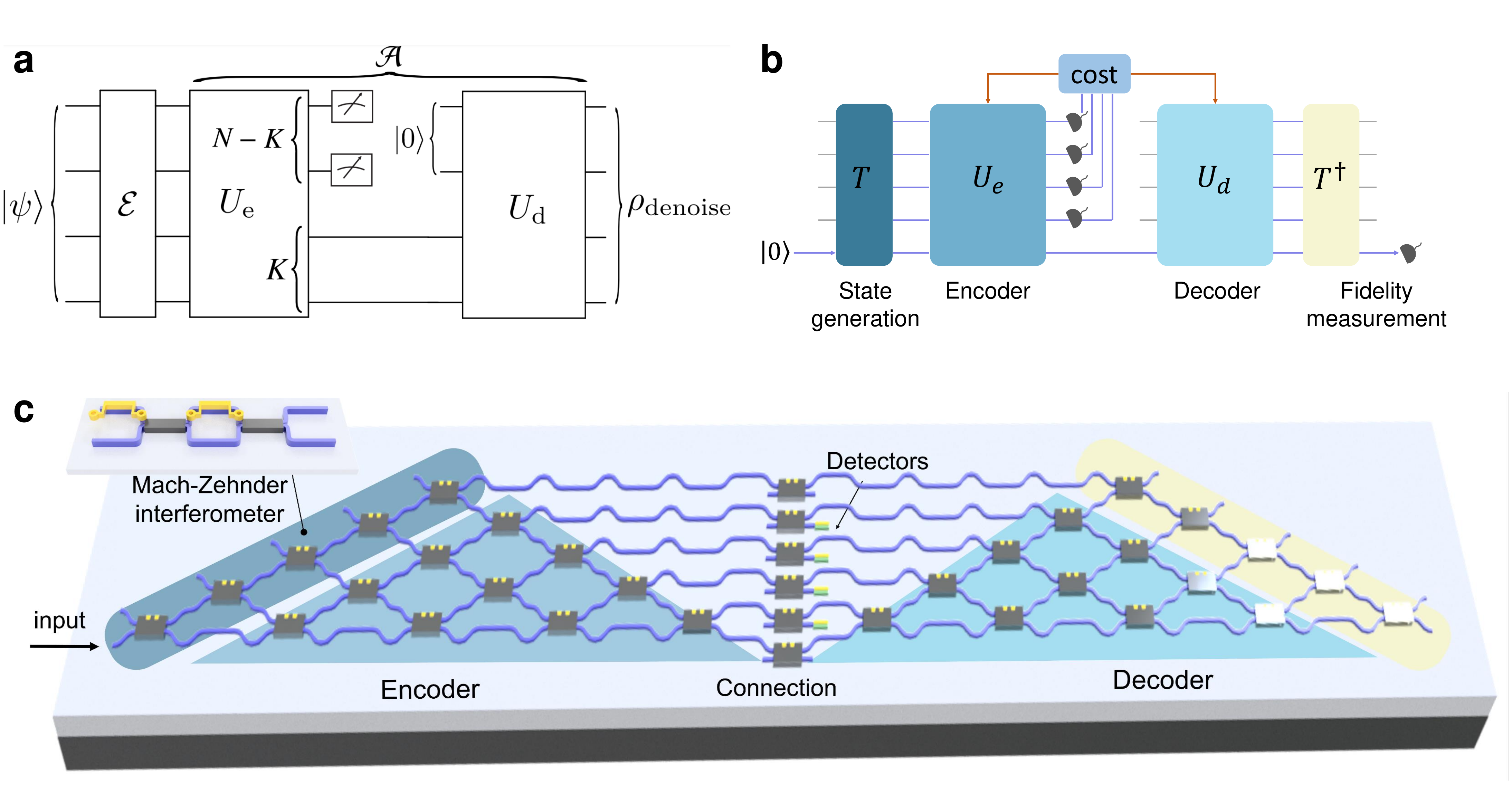}%
}
\caption{\textbf{Overview of denoising protocol and experimental implementation.} Schematic illustration of the autoencoder implemented on an integrated photonic chip for denoising quantum states. (a) Setup of the denoiser quantum autoencoder (DQA). An $N$-dimensional quantum state $\ket{\psi}$ is subject to noise channel $\mathcal{E}$. We denoise the state by encoding into the latent $K$-dimensional subspace with encoder $U_\text{e}$ and projecting out the remaining $N-K$ modes. The denoised state is constructed by decoder $U_\text{d}$. (b) The architecture of the autoencoder, which consists of the state generation unitary $T$, encoder $U_e$, decoder $U_d$, and $T^\dagger$ for measuring fidelity. Noise channels can be implemented by probabilistically choosing the state generation unitary $T$. (c) Design of the integrated photonic chip with $N=5$, which comprises two 6-by-6 linear optical circuits, one for the encoder and the other for the decoder. A column of Mach-Zehnder interferometers (MZIs) connects the encoder and decoder so that each path can be chosen to enter the decoder or go directly for measurement purposes.
}
 \label{fig:sketchFull}
\end{figure*}

\section*{Theory and Design}
\textbf{Theory of Autoencoder denoiser.} A set of pure $N$-dimensional quantum states $S=\{\ket{\psi_i}\}_i$ are affected by a noise quantum channel $\mathcal{E}$.
Our goal is to reduce the effect of the noise with an autoencoder $\mathcal{A}$ such that $\mathcal{A}(\mathcal{E}(\ket{\psi}))\approx\ket{\psi} \forall \ket{\psi}\in S$.
A sketch of the denoising protocol is shown in Fig.~\ref{fig:sketchFull}(a). A noisy input state $\rho_\text{in}=\mathcal{E}(\ket{\psi})$ is transformed with the unitary encoder $U_\text{e}(\boldsymbol{\theta})$ with trainable parameters $\boldsymbol{\theta}$. The core idea of our approach is to transform the noise into a $N-K$ dimensional redundant subspace, which is removed with the projective measurement operator $P_K =I_K \oplus 0_{N-K}$, while encoding the pure quantum information into the $K$-dimensional latent subspace. 
We post-select instances of successful projections onto $P_K$ which occur with probability
\begin{equation}
G(\rho_\text{in})=\text{tr}(P_K U_\text{e}\rho_\text{in} U_\text{e}^\dag)\,.
\end{equation}
Then, we apply the decoder unitary $U_\text{d}$ to generate the denoised state. The decoder unitary can be chosen either as the inverse of the encoder $U_\text{d}=U_\text{e}^\dagger$ or trained as $U_\text{d}(\boldsymbol{\varphi})$ with variational parameters $\boldsymbol{\varphi}$.
The final denoised state is given by
\begin{equation}
  \rho_{\text{denoise}} = \mathcal{A}(\rho_\text{in}) = \frac{1}{G(\rho_\text{in})} U_{\text{d}}P_K U_{\text{e}}\rho_\text{in} U_{\text{e}}^\dag P_K U_{\text{d}}^\dag\,.
\end{equation}
We quantify the denoiser performance via the average fidelity in respect to the ideal state $\ket{\psi}$ via
\begin{equation}\label{eq:avg_fid}
    \bar{F} = \EE{\ket{\psi}\in S}[\braket{\psi| \mathcal{A}(\mathcal{E}(\ket{\psi}))|\psi}]\,.
\end{equation}
A summary of the notations used can be found in Appendix A~\cite{supp}.
In the following, we assume that the states in $S$ span a $K$-dimensional subspace within the $N$-dimensional space of states where $N>K$. This condition ensures that, in the absence of noise, the autoencoder can achieve unit fidelity. 
We further assume that the autoencoder receives noisy states uniformly random from set $S$ where the density matrix averaged over random inputs is given by $\rho_S=\mathbb{E}_{\ket{\psi}\in S}[ \mathcal{E}(\ket{\psi})]$. 
We investigate two possible choices of decoders, which require different ways of training.
First, we assume that the decoder $U_{\text{d}}=U_{\text{e}}^\dagger$ is simply the inverse of the encoder.
This method, which we call \textit{population training}, uses the measurement probability for the cost function
\begin{equation}
C_T(\boldsymbol{\theta})=1-\text{tr}(P_K U_{\text{e}}(\boldsymbol{\theta})\rho_S U_{\text{e}}^\dagger(\boldsymbol{\theta}))\,.
\end{equation}
This cost function is maximized when incoming states have minimal probability of occupying the redundant $(N-K)$-dimensional subspace, and equivalently maximal probability in the $K$-dimensional latent space.
This can be trained in an unsupervised manner, i.e., we only require access to the noisy ensemble $\rho_S$ for training and measurements on the redundant subspace. Minimizing the cost function to the global minima yields the optimal encoder parameters $\boldsymbol{\theta}_*=\text{argmin}_{\boldsymbol{\theta}}C_T(\boldsymbol{\theta})$. The optimal encoder $U_{\text{e}}(\boldsymbol{\theta}_*)$ rotates the $K$ dominant eigenvectors of $\rho_S$ (with eigenvalues $\lambda_1 \geq \ldots \geq \lambda_K$) onto the latent subspace. Thus, the cost function is upper bounded by $C_T \le \sum_{j=1}^K \lambda_j$~\cite{cao2021noise,romero2017quantum}. It can be shown that the optimal decoder is $U_{\text{d}} = U_{\text{e}}^\dag(\boldsymbol{\theta}_*)$. This protocol has a success probability of $1 - C_T$ arising from postselection. One can think of the trained autoencoder as a projector $D_K$ onto the eigenspace spanned by the $K$ largest eigenvectors of $\rho_\text{S}$~\cite{ezzell2022quantum}.
\begin{figure*}[t]
\centering
\subfloat{%
  \includegraphics[width=0.95\linewidth]{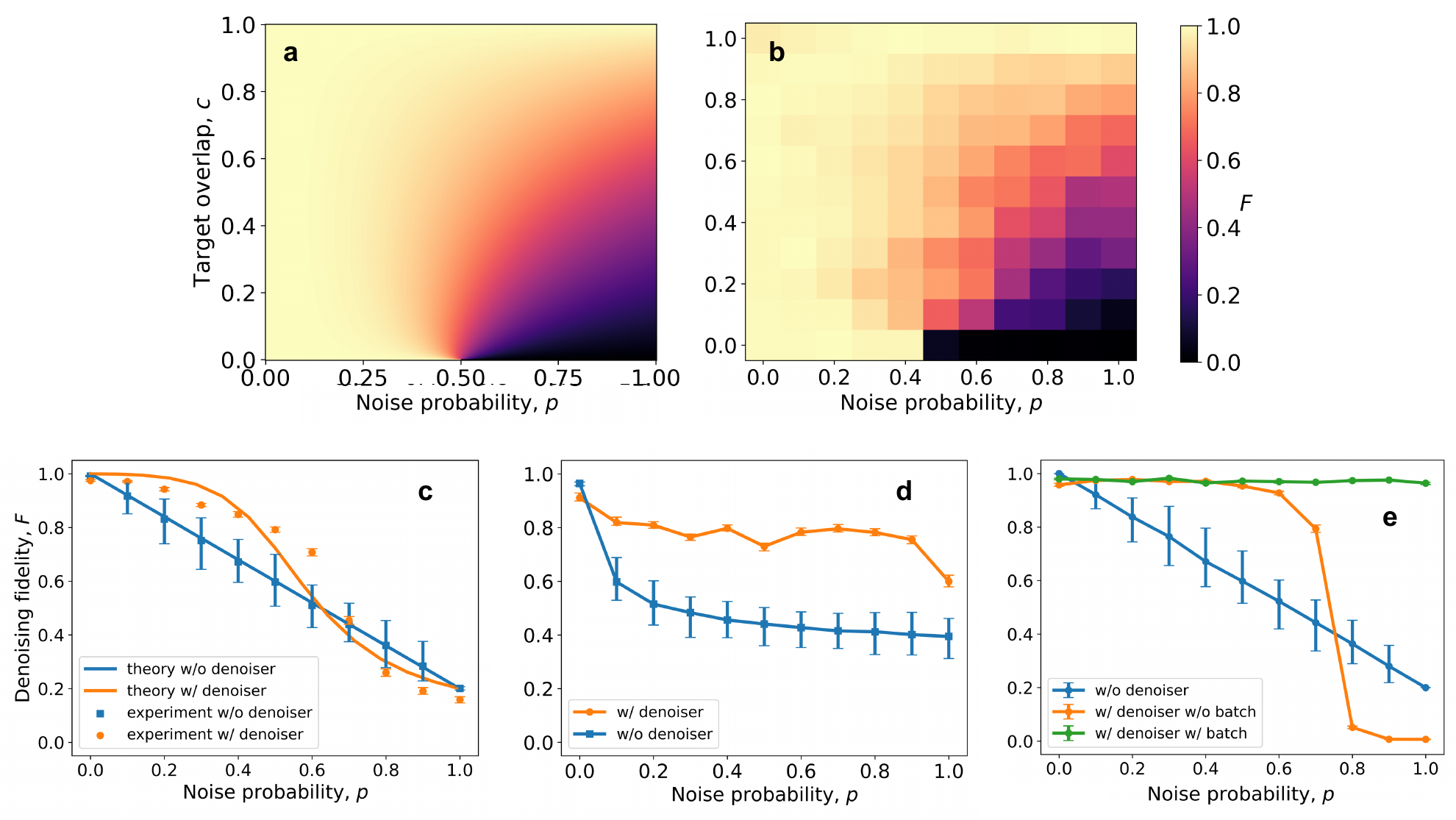}%
}
\caption{\textbf{Single state denoising, $K=1$, with population training}. (a-c) Pure state noise, with noise probability $p$ and overlap with the ideal state $c$. (a) Theoretical denoising fidelity $F$. (b) Experimental denoising fidelity for a $N=5$ qudit. (c) Denoising fidelity against $p$ for $c = 0.2$ (corresponding to the boxed values in (b)). (d) Experimental denoising fidelity for thermal noise. (e) Experimental denoising fidelity against $p$ for depolarizing noise. The theoretical value corresponds to $F = 1$ for all $p \in [0,1)$. The autoencoder has the following two training configurations: In the first configuration, a batch of five noisy states is used to train the autoencoder, ensuring the depolarizing property of the noise. In the second configuration, training is conducted without batches, and training instances are extracted individually from the set of noisy states. In cases where the sample size is not sufficiently large, this configuration might affect the depolarizing property of the noise. Population training is used for the denoiser in all cases.}
 \label{fig:singlestate}
\end{figure*}

Alternatively, we choose the decoder as $U_\text{d}(\boldsymbol{\varphi})$ to be trained separately from the encoder with its own decoder parameters $\boldsymbol{\varphi}$. In this case we perform \textit{fidelity training} to maximize the fidelity between denoised state $\mathcal{A}(\mathcal{E}(\ket{\psi}))$ and the ideal state $\ket{\psi}$
where the cost function is given by 
\begin{equation}C_F(\boldsymbol{\theta},\boldsymbol{\varphi}) = 1 - \bar{F}(\boldsymbol{\theta},\boldsymbol{\varphi})\,
\end{equation}
 which can be measured with the SWAP test.
Fidelity training requires a priori knowledge of the ideal states used for training.

\textbf{Experimental architecture and chip design.} 
We now describe the experimental implementation of our quantum autoencoder on a photonic chip, which is illustrated in Fig~\ref{fig:sketchFull}(b,c). Our chip models a $N=5$ dimensional qudit via spatial modes. The chip consists of 5 stages realized by a network of parameterized Mach-Zehnder interferometers (MZIs) and detectors. First, $N=5$ dimensional input states states are generated by the state generation stage. During this stage, we can also implement noise channels by probabilistic choosing the state generation unitary.
Then, the encoder stage realizes arbitrary unitaries with controllable circuit parameters $\boldsymbol{\theta}$. Next, up to $K\le4$ detectors realize the projective measurement to remove noise. The following decoder stage realizes arbitrary unitaries with controllable parameter $\boldsymbol{\varphi}$. To validate the denoised state, fidelity of the denoised state $\rho_\text{denoise}$ is measured by the compute-uncompute method. Given the noise-free state $\ket{\psi}=T\ket{0}$ with state generation unitary $T$, the fidelity 
is measured by applying the inverse $T^\dagger$ on $\rho_\text{denoise}$ and measuring population of the $\ket{0}$ mode~\cite{havlivcek2019supervised}.

\begin{figure*}[t]
\centering
\subfloat{%
\includegraphics[width=0.95\linewidth]{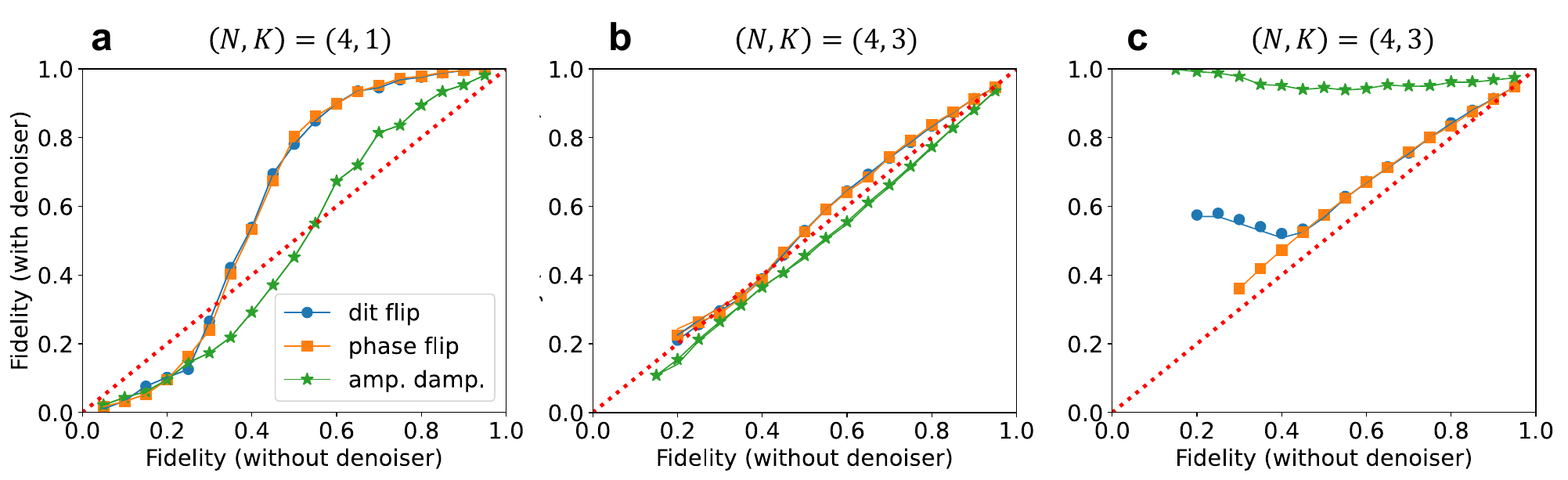}%
}
\caption{\textbf{Denoising of qudit noise channels.} Fidelity of noisy quantum states before and after denoiser. The fidelity is averaged over 
 $N=4$-dimensional states subject to dit-flip, phase flip and amplitude damping noise channels. In (a) -- (c), we sample $5 \times 10^4$ Haar random states chosen from 1000 different $K$-dimensional subspaces. Points which lie above the red dotted line indicate an improvement in fidelity by using the denoiser. We have a $K=1$ in (a), while $K=3$ in (b) and (c).  population training is used for (a) and (b), while fidelity training is used in (c). The solid lines are obtained using the analytical approximation~\eqref{eq:quenched}.}
 \label{fig:combinedKraus}
\end{figure*}

\section*{Results}
\textbf{Subspace denoising.} 
In the most general setting, the noise channel can be written as an arbitrary Kraus map $\mathcal{E}(\rho) = \sum_{n} M_n \rho M_n^\dag$ for a set of Kraus operators $M_n$ satisfying $\sum_n M_n^\dag M_n = I$. In general, obtaining an analytical expression for $\bar{F}$ is difficult. To this end, we now assume that the ideal states $S$ are uniformly sampled from a $K$-dimensional subspace via the Haar measure on the unitary group $\mathrm{U}(K)$.

This allows us to introduce a `quenched' analytical approximation~\cite{cotler2017chaos,supp}
\begin{equation}
    \bar{F}^\text{(q)} = \frac{ \sum_n (|\text{tr}(\Pi_K B M_n)|^2 + \Vert \Pi_K B M_n \Pi_K \Vert_F^2)}{(K+1) \sum_n \Vert B M_n \Pi_K \Vert_F^2}
\label{eq:quenched}
\end{equation}
where $\Pi_K$ is the projector onto the ideal subspace, $B \equiv U_{\text{d}} P_K U_{\text{e}}$, and $\Vert \cdot \Vert_F^2$ denotes the Frobenius norm. For population training, $B = D_K$, and $U_{\text{d}} = U_{\text{e}}^\dag$. 
We find that the formula agrees well with the numerical results.

A large class of realistic noise models can be obtained by setting $M_0 = \sqrt{1-p} I$, where $p$ is the noise probability. Thus, $\rho_\text{in} = (1-p)\ket{\psi}\bra{\psi}+ \sum_{n\ge1} M_n \ket{\psi}\bra{\psi}M_n^\dag$. We prove that $\Vert D_K - \Pi_K \Vert_F^2 \leq 2\sqrt{2}Kp/(1-p)$, from which we can show that the denoised state has a worst-case fidelity $\bar{F} \geq \mathbb{E}_{\ket{\psi}\in S}[\braket{\psi|\rho_\text{denoise}|\psi}] + O(p^2)$~\cite{supp}. This implies that for $p\ll1$ the denoiser is never detrimental.

We now analyze the performance of the denoiser for a more restrictive type of noise channel given by
\begin{equation}\label{eq:channelpure}
\mathcal{E}(\rho)=(1-p)\rho +p \rho_\text{noise}
\end{equation}
where the state $\rho$ is replaced with some fixed $N$-dimensional state $\rho_\text{noise}$ with probability $p$. Without the denoiser, the average fidelity of the noisy state is $\bar{F}_\text{bare}=1-p(1-c/K)$ where $c=\text{tr}(\rho_\text{noise}\Pi_K)$ 
is the overlap between the noise state and the ideal subspace. After population training, we calculate the average denoising fidelity $\bar{F}_K \equiv \mathbb{E}_S[F]$ where we find exact results for different choices of $\rho_\text{noise}$~\cite{supp}.
For depolarizing noise with $\rho_\text{noise}=I_N/N$, population training can achieve $\bar{F}_K=(1-p+p/N)/(1-p+pK/N)$  with post-selection probability $G=1-p + pK/N$.
Remarkably, for $K=1$ we find perfect denoising $F_1=1$ $\forall p\in[0,1)$.

Next, we assume a pure noise state $\rho_\text{noise}=\ket{\psi_\text{noise}}\bra{\psi_\text{noise}}$.
To quadratic order in $p$ we find~\cite{supp}
\begin{equation}
    \bar{F}_K(p) = 1 - \frac{K-1}{K} c p - \frac{K(3K-1)-1}{K}c(1-c)p^2 + O(p^3).
\label{eq:subspaceFidTaylor}
\end{equation}
The denoiser always improves the fidelity if $p \lesssim (3Kc)^{-1}$ for large $K c$~\cite{supp}. 
We now focus on the case $K=1$. Here, the denoiser suppresses noise completely to first order in $p$. By applying Theorem 1 of Ref.~\cite{koczor2021dominant}, we obtain analytically a sharp lower-bound for the fidelity of the denoised state for arbitrary pure or mixed $\rho_\text{noise}$ 
\begin{equation}
        F_{1} \geq \frac{1}{2} \left( 1 + \sqrt{1- p^2(1-p)^{-2}} \right) = 1 - \frac{1}{4}p^2 + O(p^3)
\label{eq:worstcaseFidelity}
\end{equation}
which holds for $p \leq 1/2$, whereas we get trivially $F_{1} \geq 0$ for $p > 1/2$. This is shown in Fig.~\ref{fig:singlestate}a. Notice that ${F}_1$ tends to a Heaviside step function with a sharp transition at $p=1/2$ as $c \to 0$, with the gradient diverging as $\partial_c F_1 \sim c^{-1/2}$. Intuitively, when $c = 0$, any $p > 1/2$ will cause the dominant eigenstate of $\rho$ to be orthogonal to $\ket{\psi}$ resulting in $F_1 = 0$. The experimental data in Fig.~\ref{fig:singlestate}b shows good agreement with the theoretical predictions. In particular, we plot the case of $c = 0.2$ in Fig.~\ref{fig:singlestate}c. The denoiser suppresses the noise up to linear order in $p$. Similar results are observed when using random pure states $\ket{\psi_{\text{noise}}}$ drawn from the Haar measure~\cite{supp}. For a fixed overlap $c$, the worst-case denoising fidelity corresponds to a pure $\rho_{\text{noise}}$~\cite{supp}.

Next, we experimentally demonstrate the ability of the denoiser to reduce thermal noise commonly encountered in experiments. The effects of thermal noise is modeled by adding a Gaussian random phase shift to the modes with zero mean and variance $\sigma^2$, depicted in Fig.~\ref{fig:singlestate}d. As shown in Fig.~\ref{fig:singlestate}e, the fidelity against $\ket{\psi}$ without any denoising decreases at higher variance of the noise. The denoiser improves the fidelity, demonstrating a protection against thermal noise. For the depolarizing noise we can perfectly remove the noise and achieve $F_1 = 1$ for all $p \in [0,1)$ with success probability $1-p(1-1/N)$.
Next, we consider dit-flip, phase-flip and amplitude damping channels~\cite{fonseca2019high} in Fig.~\ref{fig:combinedKraus}.
For sufficiently low noise probability $p \lesssim 0.5$, the denoiser substantially improves fidelity. The denoiser performs best for dit-flip and phase flip channels. We find only minor improvement for amplitude damping channel as it is a non-unital noise model where the steady state is pure and has a large coherent noise contribution which is hard to denoise with population training.

\textbf{Fidelity training.} 
Now, we train encoder $U(\boldsymbol{\theta})$ and decoder $U_\text{d}(\boldsymbol{\varphi})$ with separate parameters $\boldsymbol{\theta}$, $\boldsymbol{\varphi}$ by maximizing the fidelity between ideal input state and denoised state. 
In contrast to population training, fidelity training has access to the ideal state and thus can correct for coherent errors. Comparing Fig.~\ref{fig:combinedKraus}(b) and~\ref{fig:combinedKraus}(c), we see that fidelity training performs better than population training, particularly for amplitude damping noise.
We prove that, for $N\ge 2K$ and noise channel
\begin{equation}
    \mathcal{E}(\rho)=(1-p)V\rho V^\dagger+p\ket{\psi_\text{noise}}\bra{\psi_\text{noise}}
\end{equation}
with pure ideal states $\rho$, arbitrary $V$ and $\ket{\psi_\text{noise}}$, we can always find a perfect denoiser with $\bar{F}_K=1$ (see Appendix C~\cite{supp}).
\begin{figure}
\centering
\subfloat{%
\includegraphics[width=0.9\linewidth]{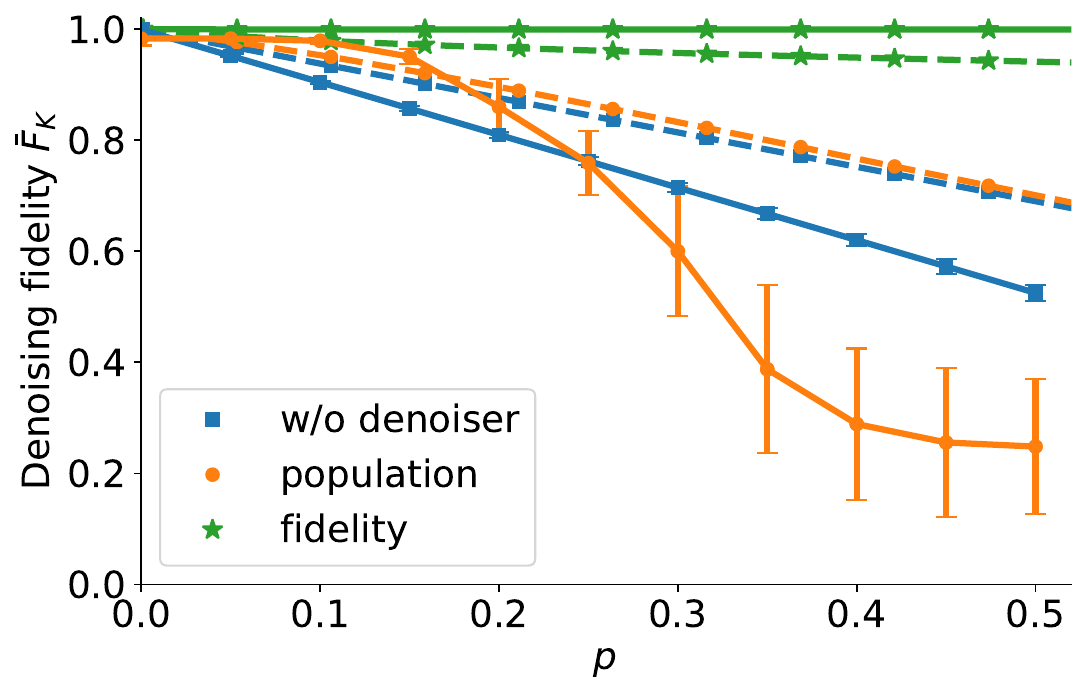}%
}
	\caption{\textbf{Subspace denoising with population and fidelity training.} Denoising fidelity, averaged over $200$ Haar random samples of the ideal subspace ($N = 3, K =2$). Solid and dashed lines represent data for $c = 0.1$ and $0.8$ respectively. The labels `population' and `fidelity' refer to the denoiser obtained from population and fidelity training respectively. Data for bare fidelity and population training with $c = 0.1$ is obtained from experiment, while the rest are obtained from numerical simulation.}
 \label{fig:subspacePure}
\end{figure}
Fig.~\ref{fig:subspacePure} shows the performance of the denoiser with both training methods, for $N = 3$ and $K = 2$. The output fidelity is averaged over 100 Haar random samples of the ideal subspace. Results for fidelity training are obtained from numerical simulations. When the noise state $\ket{\psi_{\text{noise}}}$ has little overlap with the ideal subspace $(c = 0.1)$, population training can improve the fidelity significantly for sufficiently small $p$ from \eqref{eq:subspaceFidTaylor}, but has a detrimental effect at large $p$. On the other hand, the fidelity-trained denoiser improves the output fidelity to near unity for all $p \in [0,1)$. Intuitively, the fidelity-trained denoiser is able to correct for coherent error within the ideal subspace, whereas population training can at best only project the noisy states onto the ideal subspace (or close to it) agnostic of coherent errors. This accounts for the large discrepancy in performance. To further cement this argument, we also consider a noise state which has significant overlap with the ideal subspace ($c = 0.8$). In this case, population training is essentially ineffective, whereas fidelity training can achieve a denoising fidelity of $> 0.9$ even for $p$ close to 1.

\textbf{Magic state distillation.} Magic state distillation (MSD) is an algorithm to obtain a low-noise magic state from multiple copies of noisy states. The extension of MSD to qutrits was first proposed by Anwar et al.~\cite{Anwar2012Qutrit}. However, this scheme is extremely costly due to the low success probability for each iteration of the protocol. We propose our denoiser as a pre-processing step to drastically reduce the cost of MSD. 
\begin{figure}
\centering
\subfloat{%
\includegraphics[width=0.99\linewidth]{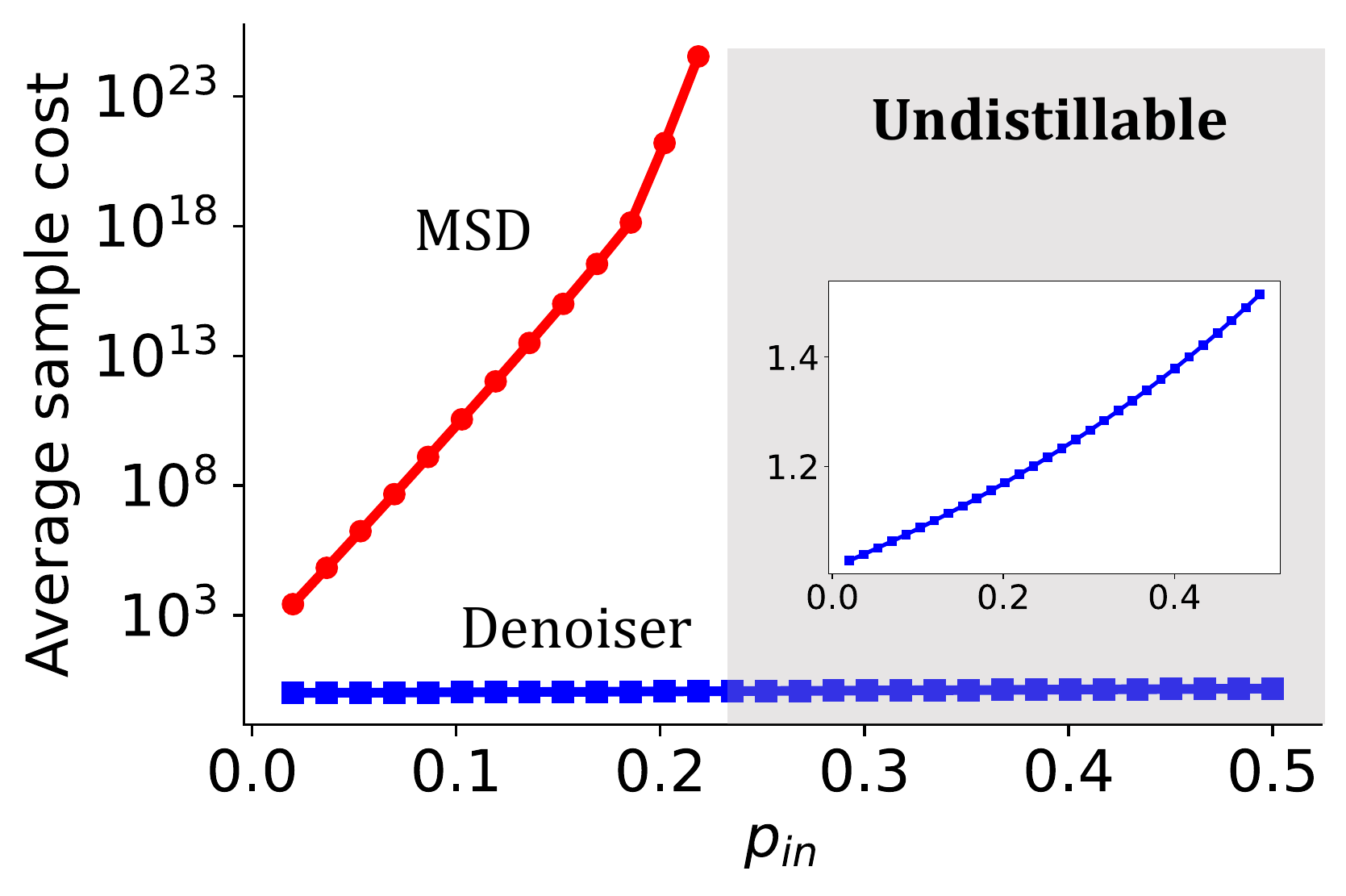}%
}
	\caption{\textbf{Denoising for magic state distillation of qutrits.} Average number of noisy magic states required to obtain a target fidelity of at least $1-2p_{\text{AE}}/3 \approx 0.987$ using MSD and the autoencoder denoiser. $p_{\text{in}}$ is the depolarizing noise probability of the input magic states. The training cost on the order of $10^2$ samples serves as a constant overhead and is omitted. In the grey region $p_{\text{in}} \gtrsim 0.233$ MSD does not work. The encoder and decoder are assumed to be noisy with each subject to depolarizing probability $p_{\text{AE}} = 0.02$.
 (Inset) We magnify the average number of noisy magic states required for denoiser. }
 \label{fig:qutritMSD}
\end{figure}
We consider the magic state $\ket{H_+}$ which is the $+1$ eigenstate of the qutrit Hadamard operator~\cite{Anwar2012Qutrit}. The magic state is distilled iteratively using the five-qutrit code $[[5,1,3]]_3$, with a success probability of around $4\%$ per iteration. The input magic states are subject to depolarizing noise with probability $p_{\text{in}}$.  
Further, we assume that the autoencoder operations itself are affected by depolarizing noise with probability $p_{\text{AE}}$ for each encoder and decoder unitary. From our experimental data, we estimate $p_{\text{AE}} = 0.02$. We compare the average number of copies of noisy magic states $N_{\text{copies}}$ needed to obtain a target fidelity of $1 - 2p_{\text{AE}}/3$, which is the fidelity attainable from our denoiser. We assume that the autoencoder has been already trained using population training with $(N,K) = (3,1)$. 
For our denoiser, $N_{\text{copies}}$ can be analytically calculated~\cite{supp} and is upper bounded by $N_{\text{copies}}\leq N = 3$. On the other hand, the sample cost for MSD is many orders of magnitude greater, as illustrated in Fig.~\ref{fig:qutritMSD}. The sample cost grows exponentially for small $p_{\text{in}}$, and diverges at the threshold $p_{\text{in}} \approx 0.233$~\cite{Anwar2012Qutrit}. Beyond the threshold noise, the magic state is undistillable while the denoiser still works efficiently. Thus, we envision that the denoiser can serve as pre-processing step to clean up noisy magic states, followed by a few iterations of MSD to reach the desired target fidelity.

\textbf{Quantum state cooling.} 
Another application of the denoiser is to cool an $N$-dimensional thermal state to the ground state~\cite{cotler2019quantum}. 
We choose the noisy input state to be a Gibbs state $\rho=\exp(-\beta H)/\mathcal{Z}$ with Hamiltonian $H$, partition function $\mathcal{Z} = \text{Tr}(\exp(-\beta H))$ and inverse temperature $\beta=1/(k_\text{B}T)$. 
We can write the state as $\rho=\mathcal{Z}^{-1}\sum_{n}\exp(-\beta E_n)\ket{\varphi_n}\bra{\varphi_n}$ with eigenenergies $E_1 \leq \dots \leq E_{N}$ and eigenstates $\ket{\varphi_n}$ of $H$. The ground state has the smallest eigenenergy $E_1$ and thus largest eigenvalue of $\rho$. Thus, population training with $K=1$ extracts the ground state $\ket{\varphi_1}$ for any $\beta$ with $F=1$ fidelity and post-selection probability $\exp(-\beta E_1)/\mathcal{Z}$.

\section*{Discussion}
We demonstrate quantum autoencoders to denoise quantum states with rigorous performance guarantees.  Our experimental demonstration on a photonic chip delivers a substantial improvement in output fidelity across a diverse range of noise channels. We propose two different variants of denoisers: Fidelity training requires noise-free reference states for training and shows exceptional performance for all considered noise models. 
In contrast, population training is only trained on the noisy input states by optimizing the probability of measuring redundant modes.
Population training shows exceptional performance in reducing incoherent errors with rigorous guarantees on the fidelity of the denoised states. For example, we reach unit fidelity for depolarizing noise and a one-dimensional subspace. The simple training protocol does not consume the denoised state such that the autoencoder can be trained online, i.e. while the autoencoder is actively denoising states. This feature could be used for adaptive online learning of the denoiser in dynamic noise environments.

Our denoiser can drastically reduce the cost of magic state distillation, a key bottleneck of fault-tolerant quantum computing, by several orders of magnitudes.
We can also cool thermal quantum states to the ground state by projecting out thermal excitations. Furthermore, our protocol could be integrated with error mitigation techniques~\cite{cai2020quantum,endo2018practical} and classical shadow tomography~\cite{huang2020predicting,koh2022classical} to enhance capabilities and reduce resource requirements, opening up new avenues for developing quantum technologies.

\section*{Methods}
\noindent\textbf{Chip fabrication}. The entire autoencoder network is manufactured on the silicon-on-insulator (SOI) platform, featuring a 220-nm-thick silicon top layer and a 2-$\mu$m-thick buried oxide layer. A thin layer of titanium nitride (TiN) is deposited as the resistive layer for heating elements. A thin aluminum film is patterned to realize the electrical connection for the heaters. Isolation trenches are etched in the SiO$_2$ top cladding and Si substrate.

\noindent\textbf{Training on-chip}. During on-chip training, we utilize coherent states as inputs to the autoencoder, leveraging their ability to achieve effects similar to those of single photons. This allows us to rapidly obtain the chip's configuration parameters. The trained autoencoder is able to denoise single photon states subject to the same noise channel. Our autoencoder is beneficial for single photon states as input, as here we can perform post-selection to denoise the state.

\noindent\textbf{Noise channel implementation}.
In the experiment, we realize noise channel $\mathcal{E}$ acting on an ideal input state $\ket{\psi}=T\ket{0}$ in the state generation stage. Here, the noisy input state $\rho_\text{noisy}=\mathcal{E}(T\ket{0})=\sum_k p_k V_k \rho V_k^\dagger$ is implemented via a probabilistic mixture of unitaries, where unitaries $V_k$ are chosen with probability $p_k$.

\noindent\textbf{Numerical simulation of the denoiser.} The optimal encoder $U_e$ for population training is equivalent to any unitary which rotates the $K$-dimensional ideal subspace onto the $K$-dominant eigenspace of the noisy input state, which we can compute numerically.  For fidelity training, the cost function is minimized by numerical optimization of the autoencoder parameters. We note that the choice of optimization algorithm has no significant impact on the denoising fidelity.

\let\oldaddcontentsline\addcontentsline
\renewcommand{\addcontentsline}[3]{}

\begin{acknowledgements}
This work is supported by a Samsung GRC project and the UKRI EPSRC grants EP/W032643/1 and EP/Y004752/1. The authors thank John Preskill, Hsin-Yuan Huang and Jielun Chen for insightful discussions.
\end{acknowledgements}

\bibliography{denoiseAutoencoder}

\let\addcontentsline\oldaddcontentsline

\clearpage
\onecolumngrid
\appendix

\setcounter{secnumdepth}{2}
\setcounter{equation}{0}
\setcounter{figure}{0}
\renewcommand{\thetable}{S\arabic{table}}
\renewcommand{\theequation}{S\arabic{equation}}
\renewcommand{\thefigure}{S\arabic{figure}}

\clearpage
\begin{center}
	\textbf{\large Appendix}
\end{center}
\setcounter{equation}{0}
\setcounter{figure}{0}
\setcounter{table}{0}
\makeatletter
\renewcommand{\theequation}{S\arabic{equation}}
\renewcommand{\thefigure}{S\arabic{figure}}
\renewcommand{\bibnumfmt}[1]{[S#1]}
We provide additional technical details and data supporting the claims in the main text.

\section{Notation and symbols}

\begin{table}[htbp]\centering
\begin{tabular}{ |l|l| } 
\hline
\textbf{Symbol} & \textbf{Name} \\
\hline
$N$ & Dimension of Hilbert space\\
$K$ & Dimension of projected subspace of autoencoder\\
$\ket{\psi}\in S$ & Set of ideal states\\
$\rho_\text{in}=\mathcal{E}(\ket{\psi})$& Noisy input state\\
  $\rho_{\text{in}}$ &  Noisy input state to autoencoder\\
          $\ket{\psi_\text{noise}}$ &  Noise perturbation\\
    $\rho_{\text{denoise}}$ &  Denoised state after autoencoder\\
$\rho_S$& Ensemble of noisy input states\\
$U_\text{e}$ & Encoder unitary\\
$U_\text{d}$ & Decoder unitary\\
$\boldsymbol{\theta}$ & Encoder parameters\\
$\boldsymbol{\varphi}$ & Decoder parameters\\
$C$ & Cost function\\
$P_K$ & Projector onto $K$ latent modes of autoencoder\\
$\Pi_K$ & Projector onto $K$-dimensional ideal subspace\\
$D_K$ & Projector onto $K$-dominant eigenspace of $\rho_S$\\
$\ket{\varphi_j}$ & Eigenvectors of $\rho_S$ with eigenvalue $\lambda_j$, in the order $\lambda_1 \geq \ldots \geq \lambda_N$ \\
$\ket{\phi_j}$ & Basis vectors of ideal subspace, $1 \leq j \leq K$\\
$\ket{\phi_j^\perp}$ & Basis vectors of orthogonal complement to ideal subspace, $1 \leq j \leq N-K$\\
 $\ket{j}$ & Computational basis vectors, $0 \leq j \leq N-1$\\
$M_n$ & Kraus operators\\
$F$ & Fidelity of a  denoised state\\
$\bar{F}$ & Ensemble average of $F$\\
$\bar{F}^\text{(q)}$ & Quenched approximation of $\bar{F}$\\
$c$ & Overlap between noise state and ideal subspace, $c = \text{Tr}(\Pi_K \rho_{\text{noise}})$\\
\hline
\end{tabular}
\caption{Definitions of symbols.}
\label{tab:definitions}
\end{table}

\section{Population training}
The cost function for population training is the population of the $N-K$ redundant modes which is to be minimized. This is equivalent to maximizing the population in the $K$ latent modes. Diagonalizing the noisy ensemble density matrix $\rho_S$, it is easy to see that the optimal encoder $U_{\text{e}}$ performs a rotation from the $K$-dominant eigenspace of $\rho_S$ to the $K$-dimensional latent subspace. The population of the latent modes is therefore the sum of the $K$-dominant eigenvalues, which is the success probability of the protocol. After projecting out the redundant modes and re-initializing them in the vacuum state, the optimal decoder $U_{\text{d}}$ is simply the inverse of the encoder, i.e., a rotation from the latent subspace back to the $K$-dominant eigenspace. Viewed together, the trained autoencoder essentially projects the noisy state onto its $K$-dominant eigenspace. Note that the choice of decoder $U_{\text{d}} = U_{\text{e}}^\dag$ is unique, since the $U_{\text{e}}$ can contain any arbitrary rotation within the latent space, which must be corrected for in the decoding step.

\subsection{Single-state denoising}
First, let us consider a noise channel of the form
\begin{equation}\label{eq:channelmixed}
\mathcal{E}(\rho)=(1-p)\rho +p \rho_\text{noise}
\end{equation}
where $\rho_\text{noise}$ is an arbitrary noise state that perturbs the ideal state $\rho$ with probability $p$.
\subsubsection{Lower bound on fidelity}
Intuitively, when the noise probability $p$ is small, the noise state acts as a perturbation to the ideal state $\ket{\psi}$. The dominant eigenstate remains close to $\ket{\psi}$, giving a denoising fidelity near unity. More concretely, by applying Theorem 1 of Ref.~\cite{koczor2021dominant}, we can bound the denoising fidelity as
\begin{equation}
    F \geq \frac{1}{2} (1 + \sqrt{1-\delta^2}),
\end{equation}
where the lower bound is saturated by noise states of the form 
\begin{align}
    \rho_{\text{noise}} &= \mu \ket{\chi}\bra{\chi} + \sum_{j=3}^{N} d_k \ket{d_k}\bra{d_k}, \\
    \ket{\chi} &= \sqrt{\frac{1-\delta}{2}} \ket{\psi} + \sqrt{\frac{1+\delta}{2}} \ket{d_2}   \label{eq:worst_sup}
\end{align}
with $\{\ket{\psi}, \ket{d_2}, \ldots, \ket{d_N} \}$ forming an orthonormal basis, and $\delta = p \mu / (1-p)$. To get the worst-case fidelity, we choose $\mu = 1$ such that $\delta = p/(1-p)$. Physically, this means that the noise state is a pure state with support only on the ideal state $\ket{\psi}$ and an orthogonal state $\ket{d_2}$. The worst-case fidelity is thus
\begin{equation}
    F_{\text{worst}} = \frac{1}{2} \left( 1 + \sqrt{1- \left(\frac{p}{1-p}\right)^2} \right) = 1 - \frac{1}{4}p^2 + O(p^3)
\label{eq:worstcaseFidelity_sup}
\end{equation}
for $p \leq 1/2$.

\subsubsection{Pure state noise}
As shown in~\eqref{eq:worst_sup} the worst-case noise state lies in the two-dimensional subspace spanned by $\ket{\psi}$ and $\ket{\psi^\perp}$. Let us now consider the noise state to be an arbitrary density matrix in the subspace
\begin{equation}
    \rho_{\text{noise}} = \begin{pmatrix} \rho_{00} & \rho_{01} \\ \rho_{01} & \rho_{00} \end{pmatrix}
\end{equation}
where $\rho_{00}=\bra{\psi}\rho_\text{noise}\ket{\psi}$ is the overlap between $\rho_\text{noise}$ and the target state $\ket{\psi}$, and $\rho_{01}$ is the coherence of $\rho_\text{noise}$ between $\ket{\psi}$ and $\ket{\psi^\perp}$. In this basis, the noisy input state can be represented by the density matrix
\begin{equation}
\rho = \begin{pmatrix}
1-p+p \rho_{00} & p \rho_{01}\\
p \rho_{01} & p(1-\rho_{00}).
\end{pmatrix}    
\end{equation}
We note that $\rho_{01}$ is upper bounded by the pure-state limit $\vert\rho_{01}\vert^2 \leq \rho_{00}(1-\rho_{00})$. The largest eigenvalue can be found exactly:
\begin{equation}
    \lambda_1 = \frac{1}{2} \left(1 + \sqrt{(1-2p(1-\rho_{00}))^2 + 4p^2 \vert\rho_{01}\vert^2} \right)
\end{equation}
with the corresponding eigenstate
\begin{equation}
\ket{\varphi_1} = \frac{1}{\mathcal{N}} \begin{pmatrix} \lambda_1 - p(1-\rho_{00}) \\ p \rho_{01} \end{pmatrix}
\end{equation}
where $\mathcal{N} = \sqrt{ (\lambda_1 - p(1-\rho_{00}))^2 + p^2 \vert\rho_{01}\vert^2 }$ is the normalization factor. The reconstruction probability is given by $\lambda_1$, and the fidelity of the reconstructed state is
\begin{equation}
    F = |\braket{\psi|\varphi_1}|^2 = \frac{1}{\mathcal{N}^2} (\lambda_1 - p(1-\rho_{00}))^2
\label{eq:fidelity2D}
\end{equation}
In the simple case where $\rho_\text{noise}$ is a pure state within the subspace spanned by $\{ \ket{\psi}, \ket{\psi^\perp} \}$, we have $\vert\rho_{01}\vert^2 = \rho_{00}(1-\rho_{00})$, and the maximal eigenvalue of $\rho$ simplifies to $\lambda_1 = \frac{1}{2}(1+\sqrt{1-4p(1-p)(1-\rho_{00})})$ and the fidelity is
\begin{equation}
\begin{split}
    F &= \left[1 + \frac{p^2 \rho_{00} (1-\rho_{00})}{ (\lambda_1 - p(1-\rho_{00}))^2 }\right]^{-1} \\&= 1 - \rho_{00}(1-\rho_{00}) p^2 + O(p^3)
\end{split}
\label{eq:PureRank2Fidelity_sup}
\end{equation}
In the limit where $\rho_{00} \to 0$, the fidelity is either 1 (when $p < 1/2$) or 0 (when $p > 1/2$). This is intuitive because it corresponds to the case where $\rho_\text{noise} = \ket{\psi^\perp}\bra{\psi^\perp}$ is orthogonal to $\ket{\psi}$, so $\ket{\psi^\perp}$ becomes the dominant eigenstate for $p > 1/2$ thus causing a sharp transition in the fidelity. More precisely, we find that
\begin{equation}
    \left.\frac{\partial F}{\partial \rho_{00}} \right|_{p=\frac{1}{2}} = \frac{1}{4\sqrt{\rho_{00}}}
\end{equation}
which diverges as $\rho_{00} \to 0$. 
\subsection{Haar-random noise state}
Suppose the noise state is now a pure state $\ket{\psi_{\text{noise}}}$ sampled from the Haar ensemble with dimension $N$. An optimal denoiser can be found for each $\ket{\psi_{\text{noise}}}$, with the denoising fidelity from~\eqref{eq:PureRank2Fidelity_sup}. We want to know what is the average-case performance of the denoiser. This can be done by integrating the fidelity over the Haar measure,
\begin{equation}
\begin{split}
    \bar{F}_N(p) &= \int_{0}^{1} d\rho_{00} \left[1 + \frac{p^2 \rho_{00} (1-\rho_{00})}{ (p_1 - p(1-\rho_{00}))^2 }\right]^{-1} \\&\times (N-1)(1-\rho_{00})^{N-2}
\end{split}
\end{equation}
For a fixed dimension $N$, the integral can be analytically computed. As an example, for $N=2$,
\begin{equation}
    \bar{F}_2(p) = \begin{cases} \frac{6+p(5p-12)}{6(p-1)^2}, &p \leq 1/2 \\ \frac{2+p}{6p}, &p > 1/2 \end{cases}
\end{equation}
which decreases monotonically from $\bar{F}_2(0) = 0$ to $\bar{F}_2(1/2) = 5/6$ to $\bar{F}_2(1) = 1/2$. In the other extreme limit of $N \to \infty$, we get the step function
\begin{equation}
    \bar{F}_{\infty}(p) = \Theta(p) - \Theta(p-1/2).
\end{equation}
For all $N$, $\bar{F}_N(1) = 1/N$, as expected from a random reconstructed state. For $p \ll 1$, we get
\begin{equation}
    \bar{F}_N(p) = 1 - \frac{N-1}{N(N+1)} p^2 - O(p^3)
\end{equation}
from which it can be shown that $\partial_N \bar{F}_N(p) > 0$ for all $N \geq 3$, and $\bar{F}_2(p) = \bar{F}_3(p)$ to order $p^2$. This means that for a sufficiently small noise probability, the denoising fidelity will increase monotonically as the dimension $N$ increases (except from $N=2 \to N=3$). The denoiser performs better on average in higher dimensions.

To determine the regime of validity, we include the $p^3$ contribution to $\bar{F}_N(p)$ which is $-4(N-1)p^3 / (N+1)(N+2)$, and demanding that this is much smaller than the $p^2$ term. This gives us the regime
\begin{equation}
p \ll \frac{N+2}{4N} \sim \frac{1}{4}.
\end{equation}
\begin{figure}
\centering
\subfloat{%
\includegraphics[width=0.5\linewidth]{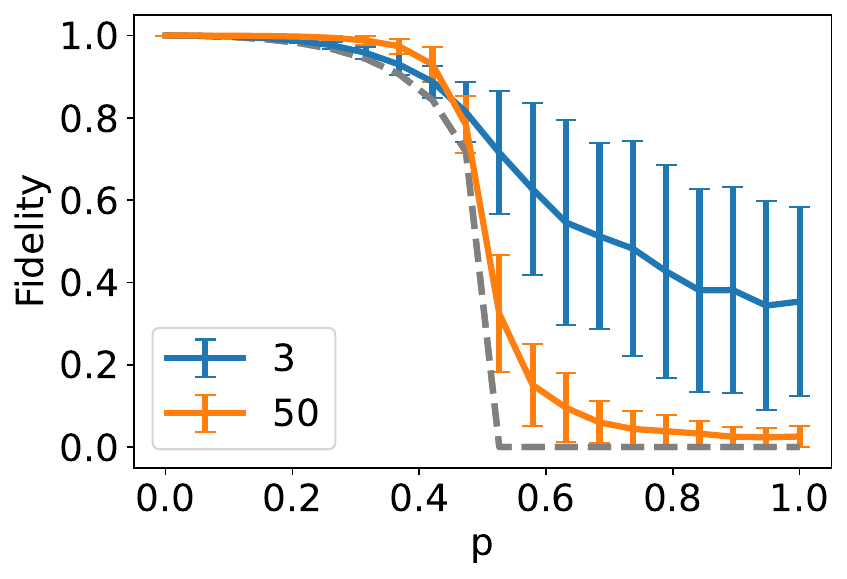}%
}
	\caption{Denoising fidelity for Haar-random noise state, with 200 instances. The solid lines represent the average fidelity $\bar{F}_N(p)$ for $N=3$ (blue) and $N=50$ (orange), while the error bars represent the standard deviation in the fidelity. The dashed line is the worst-case fidelity which provides a lower bound.}
\label{fig:fidelityHaarNoise}
\end{figure}

\subsubsection{Qudit noise channels}
\begin{figure}
\centering
\subfloat{%
\includegraphics[width=0.8\linewidth]{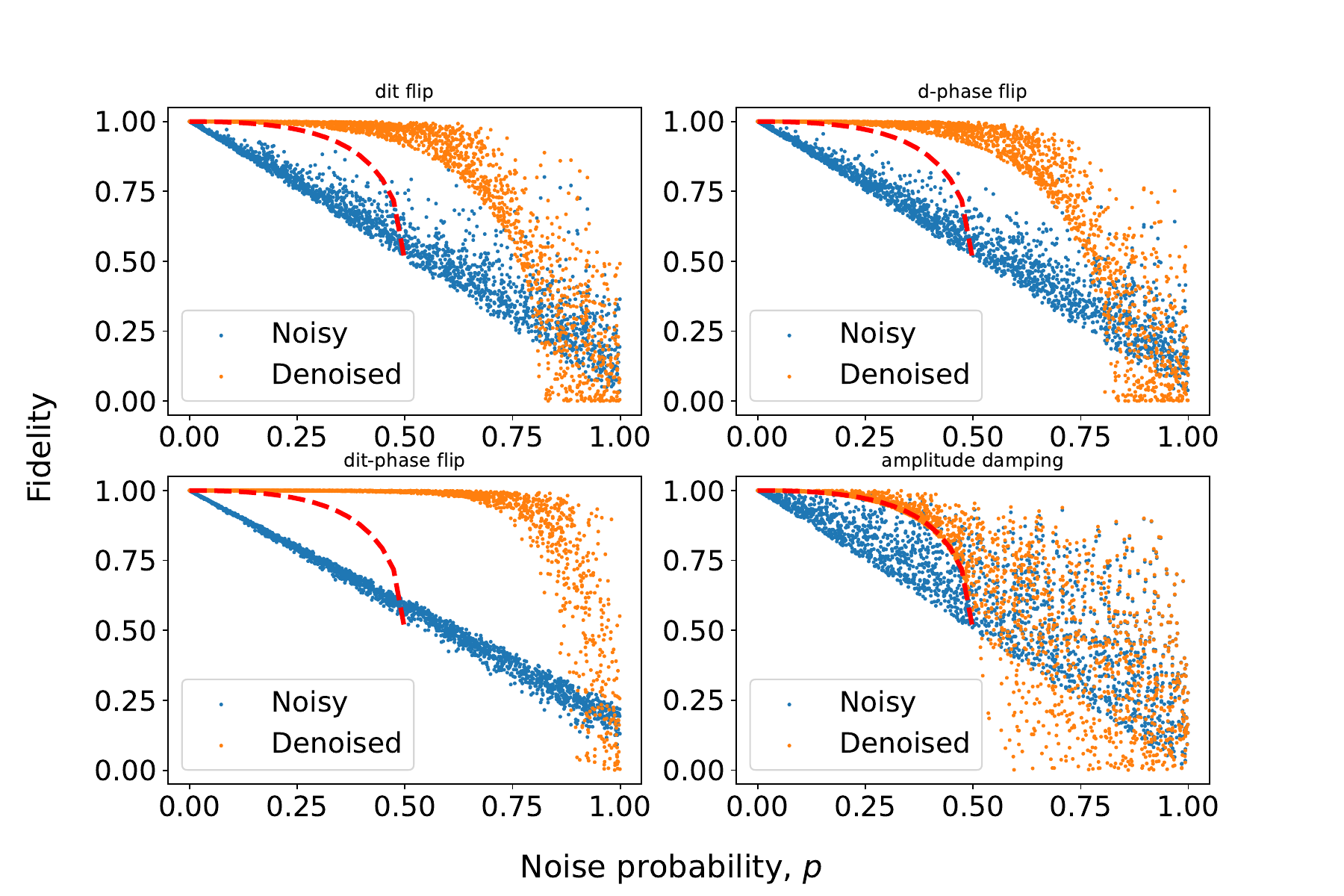}%
}
	\caption{Fidelity for $2000$ Haar-random ideal states with dimension $N = 5$ subject to dit flip, phase flip, dit-phase flip and amplitude damping channels. The red dashed line shows the worst-case fidelity for the denoised state.}
\label{fig:Kraus_N=5_K=1}
\end{figure}

Here, we consider various common qudit noise channels. Let us write the unitary Weyl operators as
\begin{equation}
W_{mn} = \sum_{j=0}^{N-1} \omega_{N}^{jm} \ket{j} \bra{j \oplus n},
\end{equation}
where $\omega_N = \exp(2\pi i /N)$, and $\oplus$ denotes addition modulo $N$. The dit-flip, phase-flip, and dit-phase flip channels can be expressed using the Weyl operators~\cite{fonseca2019high}. The Kraus operators for these noise channels are 
\begin{equation}
\begin{split}
    \text{(Dit flip)} \quad E_{00} &= \sqrt{1-p} I, \\ E_{0j} &= \sqrt{\frac{p}{N-1}} W_{0j}, \, j = 1,\ldots,N-1
\end{split}
\end{equation}
\begin{equation}
\begin{split}
    \text{(Phase flip)} \quad E_{00} &= \sqrt{1-p} I, \\ E_{j0} &= \sqrt{\frac{p}{N-1}} W_{j0}, \, j = 1,\ldots,N-1
\end{split}
\end{equation}
\begin{equation}
\begin{split}
    \text{(Dit-phase flip)} \quad E_{00} &= \sqrt{1-p} I, \\ E_{mn} &= \frac{\sqrt{p}}{N-1} W_{mn}, \, m,n = 1,\ldots,N-1
\end{split}
\end{equation}
The amplitude-damping channel, on the other hand, cannot be expressed in terms of the Weyl operators. Its Kraus operators are
\begin{equation}
\begin{split}
    \text{(Amplitude-damping)} \quad E_{0} &= \ket{0}\bra{0} + \sqrt{1-p} \sum_{j=1}^{N-1} \ket{j}\bra{j} \\ E_j &= \sqrt{p} \ket{0}\bra{j}, \, j=1,\ldots,N-1
\end{split}
\end{equation}
The denoiser performance against such noise are plotted in Fig.~\ref{fig:Kraus_N=5_K=1} for $N = 5$, for Haar-random ideal states. The denoiser performs well for dit flips, phase flips, and dit-phase flips, but not as well as for amplitude damping.

\subsection{Lower bound on population training performance}

We assume an $N$-dimensional noise channel
\begin{equation}
    \mathcal{E}(\rho)=\sum_{n} M_n \rho M_n^\dagger 
\end{equation}
with Kraus map $M_0 = \sqrt{1-p} I$ and arbitrary additional Kraus maps $M_n$, $n>0$, where $p$ is the noise probability and $\sum_n M_n^\dagger M_n= I_N$. Thus, $\rho_\text{in} = (1-p)\ket{\psi}\bra{\psi}+ \sum_{n\ge1} M_n \ket{\psi}\bra{\psi}M_n^\dag$. 
We now have ideal states randomly sampled from a $K$-dimensional subspace.
By standard integration over the ensemble, one can see that the ensemble of noisy input states in the $K$-dimensional ideal space is given by the density matrix
\begin{equation}
    \rho_S =\mathbb{E}_{\ket{\psi} \in S} [\mathcal{E}(\ket{\psi})]= (1-p) \frac{\Pi_K}{K} + p \rho_{\text{noise}}\,,
\end{equation}
where $\rho_{\text{noise}}=\frac{1}{K}\sum_{n\ge1}M_n \Pi_K M_n^\dagger$.
The ensemble $\rho_S$ has $N$ eigenvalues $\lambda_1 \geq \ldots \geq \lambda_N$ whereas $\rho_{\text{noise}}$ has eigenvalues $\nu_1 \geq \ldots \geq \nu_N$. Applying Weyl's inequality, we have
\begin{equation}
    \frac{1-p}{K} + p\nu_N \leq \lambda_i \leq \frac{1-p}{K} + p\nu_1, \quad i = 1,\ldots,K
\end{equation}
and
\begin{equation}
    p\nu_N \leq \lambda_i \leq p\nu_1, \quad i = K+1,\ldots,N.
\end{equation}
Using the Davis-Kahan $\sin \theta$ theorem~\cite{davis1970rotation} with $\delta \equiv \frac{1-p}{K} - p\nu_1$, we can upper-bound the distance between the projector onto the subspace of ideal states $\Pi_K$ and the projector $D_K$ onto the $K$ eigenvectors with largest eigenvalues of noisy states $\rho_S$ (measured via the Frobenius norm):
\begin{equation}
    \Vert \Pi_K - D_K \Vert_F \leq \frac{\sqrt{2} p}{\delta} \Vert\rho_{\text{noise}}\Vert_F.
\label{eq:daviskahan}
\end{equation}
Let us now write $\rho_{\text{noise}} = \mathbb{E}_{\ket{\psi} \in S} [\sum_{n=1} M_n \ket{\psi}\bra{\psi} M_n^\dag]$. The average fidelity after denoising is
\begin{equation}
    \bar{F} = \mathbb{E}_{\ket{\psi} \in S}\left[ \frac{(1-p)\braket{\psi|D_K|\psi}^2 + \sum_{n=1} |\braket{\psi|D_K M_n|\psi}|^2}{(1-p) \braket{\psi|D_K|\psi} + \sum_{n=1}\braket{\psi|M_n^\dag D_k M_n|\psi}} \right].
\end{equation}
From~\eqref{eq:daviskahan}, we can write 
\begin{equation}
    D_K = \Pi_K + p A
\end{equation}
for some traceless Hermitian operator $A$. Substituting this into the expression for $\bar{F}$, and using the fact that $\Pi_K \ket{\psi} = \ket{\psi} \, \forall \ket{\psi}\in S$, we have (to order $p$),
\begin{equation}
    \bar{F} = \mathbb{E}_{\ket{\psi} \in S}\left[ \frac{(1-p)(1+2p\braket{\psi|A|\psi}) +  \sum_{n=1} |\braket{\psi|M_n|\psi} + p\braket{\psi|A M_n|\psi}|^2}{(1-p) (1+p\braket{\psi|A|\psi}) + \sum_{n=1}(\braket{\psi|M_n^\dag \Pi_k M_n|\psi} + p\braket{\psi|M_n^\dag A M_n|\psi}) } \right] + \mathcal{O}(p^2).
\end{equation}
We now assume that $\Vert M_n \Vert \sim \mathcal{O}(\sqrt{p})$, giving
\begin{equation}
\bar{F} = \mathbb{E}_{\ket{\psi} \in S}\left[ \frac{(1-p)(1+2p\braket{\psi|A|\psi}) +  \sum_{n=1} |\braket{\psi|M_n|\psi}|^2}{(1-p) (1+p\braket{\psi|A|\psi}) + \sum_{n=1}(\braket{\psi|M_n^\dag \Pi_k M_n|\psi}) } \right] + \mathcal{O}(p^2)
\end{equation}
Since $\sum_{n=1} \braket{\psi|M_n^\dag \Pi_k M_n |\psi} \leq p$,
\begin{equation}
\begin{split}
\bar{F} &\geq \mathbb{E}_{\ket{\psi} \in S}\left[ \frac{(1-p)(1+2p\braket{\psi|A|\psi}) +  \sum_{n=1} |\braket{\psi|M_n|\psi}|^2}{1+p\braket{\psi|A|\psi}} \right] + \mathcal{O}(p^2) \\&= \bar{F}_{\text{bare}} + p \mathbb{E}_{\ket{\psi} \in S}[\braket{\psi|A|\psi}] + \mathcal{O}(p^2)
\end{split}
\end{equation}
where 
\begin{equation}
    \bar{F}_{\text{bare}} = \mathbb{E}_{\ket{\psi} \in S} \left[1-p+\sum_{n=1}|\braket{\psi|M_n|\psi}|^2 \right]
\end{equation}
is the average fidelity without denoising. Since $A$ is traceless, $\braket{\psi|A|\psi}$ vanishes when averaging over the Haar-random ideal states. Hence, we have the result
\begin{equation}
    \bar{F} \geq \bar{F}_\text{bare} + \mathcal{O}(p^2).
\end{equation}
Up to $O(p^2)$, the lower bound is exactly the average fidelity of the noisy state without any denoising.

\subsection{Subspace denoising}
\subsubsection{Pure state noise}
Suppose we have an ensemble of $M$ noisy states 
\begin{equation}
    \rho^{(k)} = (1-p) \ket{\psi^{(k)}}\bra{\psi^{(k)}} + p \ket{\psi_\text{noise}}\bra{\psi_\text{noise}}
\end{equation}
where $k = 1, \ldots, M$. $\ket{\psi^{(k)}}$ lies in the $K$-dimensional ideal space, while $\ket{\psi_\text{noise}}$ is a fixed noise state in the full $N$-dimensional Hilbert space. We assume that $\ket{\psi^{(k)}}$ is sampled from the Haar distribution. A uniform mixture of $\rho^{(k)}$ gives the density matrix
\begin{equation}
    \rho_S = \frac{1}{M} \sum_{k=1}^{M} \rho^{(k)} \approx (1-p)\frac{I_K}{K} + p \ket{\psi_\text{noise}}\bra{\psi_\text{noise}}
\end{equation}
with the approximation becoming exact as the sample size $M \to \infty$. Note that while we have added $\ket{\psi_{\text{noise}}}$ incoherently to $\ket{\psi^{(k)}}$, the same density matrix can also be obtained if we treat $\ket{\psi_{\text{noise}}}$ as a coherent noise. To see this, we now write $\rho^{(k)} = \ket{\psi_{\text{in}}^{(k)}}\bra{\psi_{\text{in}}^{(k)}}$, where 
\begin{equation}
    \ket{\psi_{\text{in}}} = \sqrt{1-p} \ket{\psi^{(k)}} + \sqrt{p} \ket{\psi_{\text{noise}}}.
\end{equation}
Taking a uniform mixture, the density matrix is now
\begin{equation}
\begin{split}
    \rho_S &\approx (1-p)\frac{I_K}{K} + p \ket{\psi_\text{noise}}\bra{\psi_\text{noise}} \\&+ \sqrt{p(1-p)} \mathbb{E}\left[\ket{\psi^{(k)}}\bra{\psi_\text{noise}} + \ket{\psi_\text{noise}}\bra{\psi^{(k)}}\right]
\end{split}
\end{equation}
where $\mathbb{E}[\cdot]$ denotes an average over the ensemble of states. Assuming a Haar ensemble, the extra term vanishes on average, so the resulting density matrix is the same as that with incoherent noise.

\begin{figure}
\centering
\subfloat{%
\includegraphics[width=0.5\linewidth]{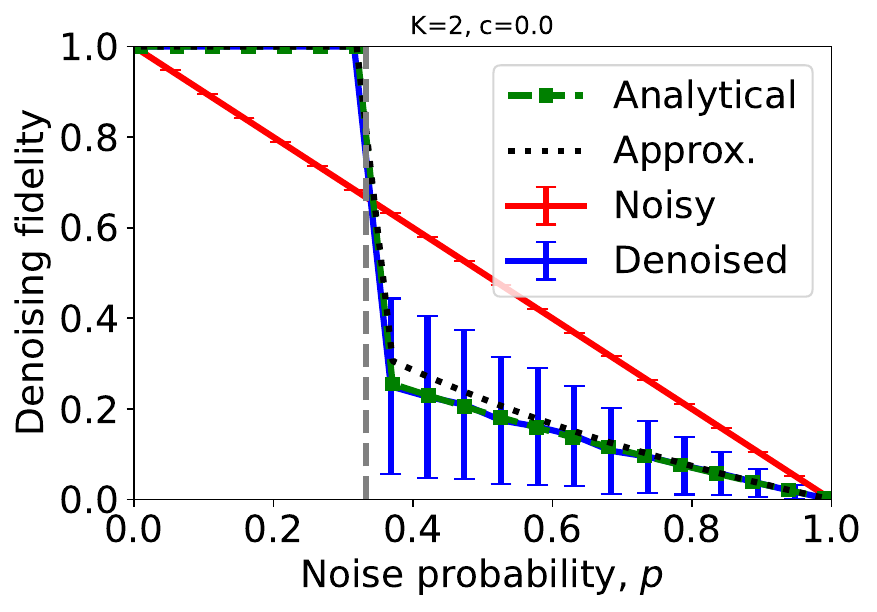}%
}
	\caption{Denoising fidelity for Haar-random states drawn from a $K$-dimensional $(K=2)$ ideal subspace, with 1000 instances. The noise state is orthogonal to the ideal subspace. The solid line represents the average fidelity $\bar{F}_K(p)$ for the denoised state (blue) and initial noisy state (red), while the error bars represent the standard deviation. The exact average fidelity is given by the green square markers while the approximate average fidelity is given by the black dotted line. The grey dashed line marks $p = 1/(K+1)$.}
	\label{fig:fidelitySubspaceC0}
\end{figure}

Let us write the noise state as
\begin{equation}
    \ket{\psi_\text{noise}} = \sqrt{c} \ket{\xi} + \sqrt{1-c} \ket{\xi^\perp} 
\end{equation}
where $\ket{\xi}$ is a basis state in the ideal subspace, and $\ket{\xi^\perp}$ is a basis state in the orthogonal complement of the ideal subspace. In an orthonormal basis containing $\ket{\xi}$ and $\ket{\xi^\perp}$, the density matrix has a block diagonal form
\begin{equation}
    \rho_S = \begin{pmatrix} \frac{1-p}{K} + p c & p\sqrt{c(1-c)} \\ p\sqrt{c(1-c)} & p(1-c)  \end{pmatrix} \oplus \frac{1-p}{K} I_{K-1} \oplus \boldsymbol{0}_{N-K-1}.
\end{equation}
The non-trivial eigenvalues and eigenvectors of $\rho_S$ are
\begin{equation}
    \lambda_{\pm} = \frac{1}{2K} \bigg[ 1+p(K-1) \pm \sqrt{1+p(-2+p+K(-2-4c(-1+p)+(2+K)p))} \bigg]
\end{equation}
\begin{equation}
    \ket{\lambda_\pm} = \frac{1}{\mathcal{N}_{\pm}} \left[ p\sqrt{c(1-c)} \ket{\xi} + \left(\lambda_\pm - \frac{1-p}{K} - p c\right) \ket{\xi^\perp} \right]
\end{equation}
with the normalization factors
\begin{equation}
    \mathcal{N}_\pm^2 = p^2 c(1-c) + \left(\lambda_\pm - \frac{1-p}{K} - p c\right)^2.
\end{equation}
The other eigenvalues are $(1-p)/K$ (with degeneracy $K-1$) and $0$ (with degeneracy $N-K-1$). It can be shown that for $0 < c < 1$, $\lambda_- < (1-p)/K < \lambda_+$, so the optimally trained denoiser projects the state onto the subspace spanned by $\ket{\lambda_+}$ and the remaining $K-1$ orthonormal basis vectors $\ket{\xi_1} \ldots \ket{\xi_{K-1}}$ in the ideal space. Denoting $\alpha \equiv |\braket{\xi|\psi}|^2$, $\beta \equiv |\braket{\xi|\lambda_+}|^2$ and $\gamma \equiv |\braket{\lambda_+|\psi_\text{noise}}|^2$, the denoising fidelity is obtained as
\begin{equation}
    F_K(p) = \frac{(1-p)(\alpha\beta + 1 - \alpha)^2 + p \alpha \beta \gamma}{(1-p)(\alpha \beta + 1 - \alpha)+p\gamma}.
\label{eq:subspaceFidelity}
\end{equation}
This is dependent on the choice of the ideal state $\ket{\psi}$. We can remove this dependence by averaging over the Haar-random states $\ket{\psi}$, which gives
\begin{equation}
    \bar{F}_K(p) = (K-1)\int_{0}^1 d\alpha (1-\alpha)^{K-2} F_K(p).
\end{equation}
In the degenerate case where $c=0$ (noise state is orthogonal to the ideal subspace), the integral can be solved exactly to give
\begin{equation}
    \lim_{c\to0}\bar{F}_K(p) = \begin{cases}1 &, p < \frac{1}{K+1} \\ \frac{K-1}{K+1}(1-p) {}_{2}\mathcal{F}_{1}(1,1;K-2;1-p)&, p \geq \frac{1}{K+1} \end{cases} 
\end{equation}
where ${}_{2}\mathcal{F}_{1}$ is the hypergeometric function. The denoiser performance transitions sharply from perfect recovery ($p < 1/(K+1)$) to becoming detrimental ($p > 1/(K+1)$), as depicted in the numerical simulations in Fig.~\ref{fig:fidelitySubspaceC0}. 

For a non-zero $c$, by performing a partial fraction expansion of $\bar{F}_K(p)$ and integrating term-by-term, the integral can be done exactly, to yield
\begin{equation}   
\begin{split}
\bar{F}_K(p) &= -\frac{1-\beta}{K} + \left(1 - \frac{p\gamma}{(1-p)(1-\beta)} \right) \\ &- \frac{p\gamma(\beta-p \beta - p\gamma}{(1-p)(1-\beta)(1-p+p\gamma)} \\ &\times {}_{2}\mathcal{F}_{1}(1,1;K;(1-p)(1-\beta)/(1-p+p\gamma)).
\end{split}
\end{equation}
To order $p^2$, we have
\begin{equation}
    \bar{F}_K(p) = 1 - \frac{K-1}{K} c p - \frac{K(3K-1)-1}{K}c(1-c)p^2 + O(p^3).
\end{equation}
As a consistency check, we set $K=1$, which recovers~\eqref{eq:PureRank2Fidelity_sup} ($c = \rho_{00}$). This means that if we want to denoise a subspace beyond just a single state, the leading-order correction to the denoising fidelity is proportional to $p$ instead of $p^2$, hence the denoiser becomes less effective. Nevertheless, we can show that for a sufficiently small $p$, using the denoiser is still advantageous. Without the denoiser, the average fidelity is
\begin{equation}
    \bar{F}_K^\prime(p) = 1 - \left(1-\frac{2c}{K(K+1)}\right)p.
\end{equation}
The denoising fidelity is higher than $\bar{F}_K^\prime$ if 
\begin{equation}
    p < \frac{K(K+1)-c(K^2+1)}{c(1-c)(K+1)(K(3K-1)-1)} \sim \frac{1}{3Kc}
\end{equation}
for large $K c$. We can see that range of $p$ for which the denoiser is useful shrinks like $1/K$, so high-dimensional subspace denoising only works for very small noise probabilities.
\begin{figure}
\centering
\subfloat{%
\includegraphics[width=0.5\linewidth]{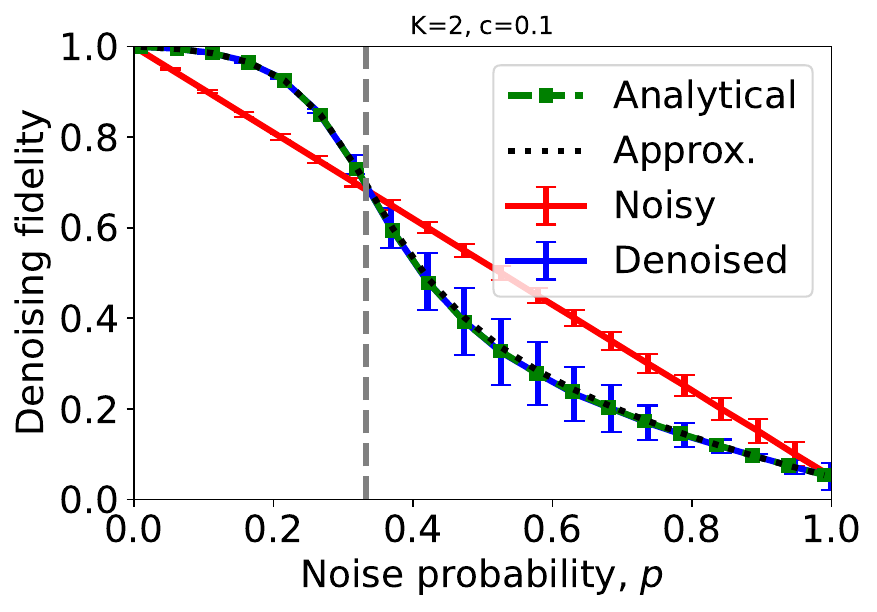}%
}
	\caption{Denoising fidelity for Haar-random states drawn from a $K$-dimensional $(K=2)$ ideal subspace, with 1000 instances. The noise state has an overlap of $c = 0.1$ with the ideal subspace. The solid line represents the average fidelity $\bar{F}_K(p)$ for the denoised state (blue) and initial noisy state (red), while the error bars represent the standard deviation. The exact average fidelity is given by the green square markers while the approximate average fidelity is given by the black dotted line. The grey dashed line marks $p = 1/(K+1)$.}
	\label{fig:fidelitySubspaceCnonzero}
\end{figure}
A more compact approximate expression can be obtained by exploiting properties of the Haar distribution, giving
\begin{equation}
\begin{split}
\bar{F}_K^{(\text{approx.})}(p) &= \bigg[\frac{1-p}{K+1} (\text{tr}(\Pi_K{D_K})^2+\text{tr}((\Pi_K{D_K})^2) )+p\text{tr}(\Pi_K{D_K}\sigma{D_K})\bigg]\bigg/ [(1-p)\text{tr}(\Pi_K{D_K})\\&+Kp\text{tr}({D_K} \sigma)], 
\end{split}
\end{equation}
where $\Pi_K$ is the projector onto the ideal subspace, and $D_K$ is the projector on the $K$-dominant eigenspace of the noisy ensemble $\rho$. An example of $K=2, c=0.1$ is plotted in Fig.~\ref{fig:fidelitySubspaceCnonzero}. We can see that the approximate formula agrees well with the exact expression for average fidelity.

\subsubsection{Quenched approximation for general channels}
We now derive the quenched approximation for the fidelity $\bar{F}^\text{(q)}$ for general quantum channels.
We assume that the pure input states are subject to general Kraus operators $M_n$ with the condition $\sum_n M_n^\dagger M_n=I$. The pure input states $\ket{\psi_j}$ transform to $\rho_j=\sum_{n} M_n \ket{\psi_j}\bra{\psi_j} M_n^\dagger$. 
Averaging over Haar-random (or at least a 2-design) input states $\ket{\psi_j}$ in the $K$-dimensional ideal subspace, the ensemble is described by the density matrix
$\rho_S=\frac{1}{K}\sum_{n}M_n \Pi_K M_n^\dagger$. Denoting the (unnormalized) optimal autoencoder operation as $B \equiv U_{\text{d}} P_K U_{\text{e}}$, for each input the output state can be written as
\begin{equation}
    \rho_{\text{out},j} = \frac{\sum_n B M_n \rho_j M_n^\dag B^\dag}{\sum_n \text{tr}(B M_n \rho_j M_n^\dag B^\dag)},
\end{equation}
and the fidelity with the ideal state $\ket{\psi_j}$ is $F_j = \braket{\psi_j\vert\rho_{\text{out},j}|\psi_j}$. By Haar integrating over the numerator and denominator separately we get an approximate expression for the average fidelity $\bar{F}^\text{(q)} \approx \bar{F} = \mathbb{E}_{j}[F_j]$,
\begin{equation}
    \bar{F}^\text{(q)} = \frac{ \sum_n (|\text{tr}(\Pi_K B M_n)|^2 + \Vert \Pi_K B M_n \Pi_K \Vert_F^2)}{(K+1) \sum_n \Vert B M_n \Pi_K \Vert_F^2}
\end{equation}
where $\Vert \cdot \, \Vert_F^2$ denotes the Frobenius norm. For population training, $B = D_K$, and $U_{\text{d}} = U_{\text{e}}^\dag$.

\subsection{Noisy autoencoder}
We have largely assumed that the denoiser itself is noiseless when analyzing its performance. However, any realistic implementation of the autoencoder will invariably contain some noise. For simplicity, we model the noise of the encoder and decoder as independent depolarizing channels with the same probability $p_{\text{AE}}$. From the experimental data, we get a crude estimate of $p_{\text{AE}} \approx 0.02$. Let us analyze the denoising of a single state $(K = 1)$ with this noise. Denoting the input noise channel as $\mathcal{E}$, the denoising probability (which is the post-selection probability) is the dominant eigenvalue of $\rho$, where
\begin{equation}
    \rho = (1-p_{\text{AE}}) \mathcal{E}(\ket{\psi}) + p_{\text{AE}} \frac{I_N}{N}.
\end{equation}
The corresponding dominant eigenstate is denoted $\ket{\chi_0}$. The output state is
\begin{equation}
    \rho_{\text{out}} = (1-p_{\text{AE}}) \ket{\chi_0}\bra{\chi_0} + p_{\text{AE}} \frac{I_N}{N},
\end{equation}
and the denoising fidelity is 
\begin{equation}
    F = (1-p_{\text{AE}})|\braket{\chi_0|\psi}|^2 + \frac{p_{\text{AE}}}{N}.
\end{equation}
In the limit $p_{\text{AE}} = 0$, we recover $\ket{\chi_0} \to \ket{\psi}$ and $F \to 1$.

If we further assume that the input noise is also an independent depolarizing noise with probability $p_{\text{in}}$, the denoising probability becomes
\begin{equation}
    p_{\text{denoise}} = (1-p_{\text{AE}})(1-p_{\text{in}}) + \frac{p_{\text{in}}+p_{\text{AE}}(1-p_{\text{in}})}{N}
\end{equation}
with denoising fidelity
\begin{equation}
    F = 1 - \frac{N-1}{N} p_{\text{AE}}.
\end{equation}
In the worst-case of $p_{\text{in}} = 1$, we need to repeat this protocol on average $N$ times to successfully denoise the state.

\section{Fidelity training}
For fidelity training, we assume to have access to noiseless ideal states during the training stage (this is not required for population training). The cost function is the average fidelity between the output states and the ideal states, which can be implemented using a SWAP test. The advantage is that it can correct for coherent noise within the ideal subspace, while population training is more suited for incoherent noise. A denoiser obtained from fidelity training always outperforms that from population training, with the trade off being the increased training cost.

\subsection{Sufficient condition for perfect denoising}
We claim that for the fixed noise channel and any fixed pure noise state $\rho_\text{noise}=\ket{\psi_\text{noise}}\bra{\psi_\text{noise}}$, we can achieve perfect denoising where the average fidelity is $1$ $\forall p,c$. This is possible if 
$N \geq 2K$. To prove this, notice that a necessary condition for perfect denoising is to losslessly compress the ideal subspace while simultaneously removing the noise, i.e. $P_K U_{\text{e}} \Pi_K U_{\text{e}}^\dag P_K = P_K$ and $P_K U_{\text{e}} \ket{\psi_\text{noise}} = 0$, where $P_K$ is the projector onto the $K$ latent modes and $\Pi_K$ is the projector onto the ideal subspace. Without loss of generality, for $c\geq0$, we can write $\ket{\psi_\text{noise}} = \sqrt{c} \ket{\phi_1} + \sqrt{1-c} \ket{\phi_1^\perp}$, $\ket{\psi_\text{noise}^\perp} = \sqrt{1-c} \ket{\phi_1} - \sqrt{c} \ket{\phi_1^\perp}$, and the ideal state $\ket{\psi} = \sum_j b_j \ket{\phi_j}$ with $\sum_j |b_j|^2 = 1$. We use two sets of basis states: $\{\ket{j}\}_{j=1}^{N+K}$ which spans the latent space, and $\{\ket{\phi_j}\}_{j=1}^N\cup\{\ket{\phi_j}^{\perp}\}_{j=1}^K$ which spans the ideal subspace. We can construct
\begin{equation}
\begin{split}
    U_1 &= \ket{N-1}\bra{\psi_{\text{noise}}} + \ket{0}\bra{\psi_{\text{noise}}^\perp} + \sum_{j=1}^{K-1} \ket{j}\bra{\phi_{j+1}} \\&+ \sum_{j=2}^{N-K} \ket{j+K-2}\bra{\phi_j^\perp}
\end{split}
\end{equation}
which rotates $\ket{\psi_{\text{noise}}}$ to one of the redundant modes, and for $N \geq 2K$ we define
\begin{equation}
\begin{split}
    U_2 &= \sum_{j=1}^{K-1} \bigg[(\sqrt{1-c}\ket{j} + \sqrt{c}\ket{j+K-1})\bra{j} \\&+ (\sqrt{c}\ket{j} - \sqrt{1-c} \ket{j+K-1})\bra{j+K-1}\bigg] + \ket{0}\bra{0} \\&+ \sum_{j=2K-1}^{N-1} \ket{j}\bra{j}
\end{split}
\end{equation}
which is required to losslessly compress the ideal subspace. The encoder unitary is thus given by $U_{\text{e}} = U_2 U_1$, with
\begin{equation}
\begin{split}
    &P_K U_{\text{e}} \ket{\psi_{\text{noise}}} = 0, \\
    &P_K U_{\text{e}} \ket{\psi} = \sqrt{1-c} \sum_{j=1}^{K} b_j \ket{j-1}.
\end{split}
\end{equation}
Normalizing the latent state $P_K U_{\text{e}} \ket{\psi}$ and applying the decoder unitary 
\begin{equation}
    U_{\text{d}} = \sum_{j=1}^{K} \ket{\phi_{j}}\bra{j-1} + \sum_{j=1}^{N-K} \ket{\phi_j^\perp}\bra{j+K-1}
\end{equation}
results in perfect denoising with success probability $1-c$. 

A dimension of $N\ge 2K$ is necessary to perfectly reconstruct the input state. This can be seen from the fact that the space of possible input states spans a $K$-dimensional subspace. Due to unitary constraints, preparing the states that can be deterministically denoised for all possible inputs requires at least a $K$-dimensional auxiliary space, thus requiring in total $2K$-dimensional $U_{\text{e}}$ and $U_{\text{d}}$ $\square$.

\section{Coherent states and single photon Fock states}

For a large class of noise models, one can train our quantum autoencoder with coherent states as input. After training this way, the trained autoencoder can successfully denoise single photon input states subject to the same noise model.

To see this, first note that transformations with linear optics over $N$ modes can be described by a $N\times N$ unitary $U$. This unitary transforms the creation operator of the $k$th mode $\hat{a}_k^\dagger$  via $\hat{a}_k^\dagger\rightarrow \sum_{\ell} U_{k\ell} \hat{a}_{\ell}^\dagger$. 
We now discuss the effect of $U$ on single photon Fock states and coherent state as input to the quantum autoencoder.

For a single photon input on the first mode, after application of $U$ we get $\ket{\psi_\text{F}}=U\hat{a}_1^\dagger\ket{\text{vac}}=\sum_{\ell} U_{1\ell}\hat{a}_\ell^\dagger\ket{\text{vac}}$, where $\ket{\text{vac}}$ is the vacuum state without photons. Measurement of the average photon number in mode $k$ gives us $\bra{\psi_\text{F}}\hat{a}_k^\dagger\hat{a}_k\ket{\psi_\text{F}}=\vert U_{1k}\vert^2$.

For a coherent state input on the first mode we have $\ket{\psi_\text{C}}=U D_1(\alpha)\ket{\text{vac}}= \prod_{\ell}D_\ell(U_{1\ell}\alpha)\ket{\text{vac}}$ where $D_\ell(\alpha)=\exp{(-\alpha \hat{a}_\ell-\alpha^\ast \hat{a}_\ell^\dagger)}$ is the displacement operator acting on mode $\ell$ and $\alpha$ the amplitude of the coherent state. The average photon number of mode $k$ is given by $\mathcal{I}_k=\bra{\psi_\text{C}}\hat{a}_k^\dagger\hat{a}_k\ket{\psi_\text{C}}=\vert \alpha\vert^2\vert U_{1k}\vert^2$. Thus, training on the population on the trash mode  for coherent states and single photon Fock states is equivalent up to a constant scaling factor. As the fidelity is measured using the photon population after the inverse of the state generation unitary, the equivalence also applies to fidelity training,

By extending above arguments to mixtures, we note that single photons and coherent state inputs also show equivalent populations under a large class of noise channels affecting the state. In particular, the equivalence holds for any noise channel $\mathcal{E}(\rho)=\sum_{k}V_k \rho V_k^\dagger$ where the Kraus operators $V_k$ can be expressed in terms of linear optical elements. 

\end{document}